\documentclass{PoS}
\usepackage{slashed}

\def\WpWm{$W^+W^-$}
\def\WpWmupjj{$W^+W^-+\!\le2$}
\def\WpWmjj{$W^+W^-\!+2$}
\def\WpWmjjj{$W^+W^-\!+3$}
\newcommand{\BlackHat}{{\sc BlackHat}}
\newcommand{\SHERPA}{{\sc SHERPA}}
\newcommand{\Powheg}{{\sc Powheg}}
\newcommand{\MCatNLO}{{\sc MC@NLO}}

\title{Di-vector Boson Production with Jets at the LHC}

\ShortTitle{Di-bosons and Jets at the LHC}

\author{\speaker{Fernando Febres Cordero}, Harald Ita \\
        Physikalisches Institut, Albert-Ludwigs-Universit\"at, Freiburg\\
        D-79104 Freibug, Germany\\
        E-mail: \email{ffebres@physik.uni-freiburg.de}}

\abstract{In this talk we present the first calculation of next-to-leading-order
QCD corrections for the production of \WpWm{} pairs in association with three jets
at the LHC. We show the observed improvement in the dependence of total and
differential cross sections on the unphysical renormalization and factorization
scales. We study the radiation pattern for configurations associated to
vector-boson fusion and the impact that the QCD corrections have on them.}

\FullConference{Loops and Legs in Quantum Field Theory\\
		24-29 April 2016\\
		Leipzig, Germany}

\begin{document}

\section{Introduction}

Signatures of di-vector boson production in association with jets are
associated to very rich phenomenology. They are key to understanding the
electroweak gauge structure, as they can be used to measure trilinear and
quartic gauge couplings and to put constraints on anomalous gauge couplings.
They appear naturally in top physics, since final-state configurations in
$t\bar t$ production contain a \WpWm{} pair and two or more jets. The Higgs
boson can decay into a vector-bosons pair, for example \WpWm{}, $ZZ$,
$\gamma\gamma$ and $Z\gamma$, and is accompanied by jets in its production
through vector boson fusion. Also in many models beyond the Standard Model (SM)
decay chains started by pairs of heavy colored particles lead to the associated
production of several leptons and jets, for which the SM production of two weak
bosons and jets are main backgrounds. 

In this talk we present details of our next-to-leading-order (NLO) QCD
study~\cite{ww3j} of the production of a \WpWm{} pair in association with
up to three jets at the LHC. Related calculations have been
published before, starting with studies of the inclusive \WpWm{} rates at leading order
(LO)~\cite{WWLO}, NLO~\cite{WWNLO} and reaching up to next-to-next-to-leading (NNLO) order
in QCD~\cite{WWNNLO}. The case of \WpWm{} hadro production in association with one jet
was studied at NLO QCD in ref.~\cite{WW1j} and with two jets
in ref.~\cite{WW2ja,WW2jb,Alwall:2014hca}. Studies of same-sign
$W$ pair production with two jets were presented in ref.~\cite{WpWp2j}.
Our calculation for \WpWmjjj{} jets is a state-of-the-art large-multiplicity NLO
QCD calculation at hadron colliders which presently reach
five~\cite{W4jBH,Z4jBH,OtherNLO2to5} or six~\cite{W5jBH} objects in the final
state.

\begin{figure*}[t]
\begin{center}
\includegraphics[clip,scale=0.35]{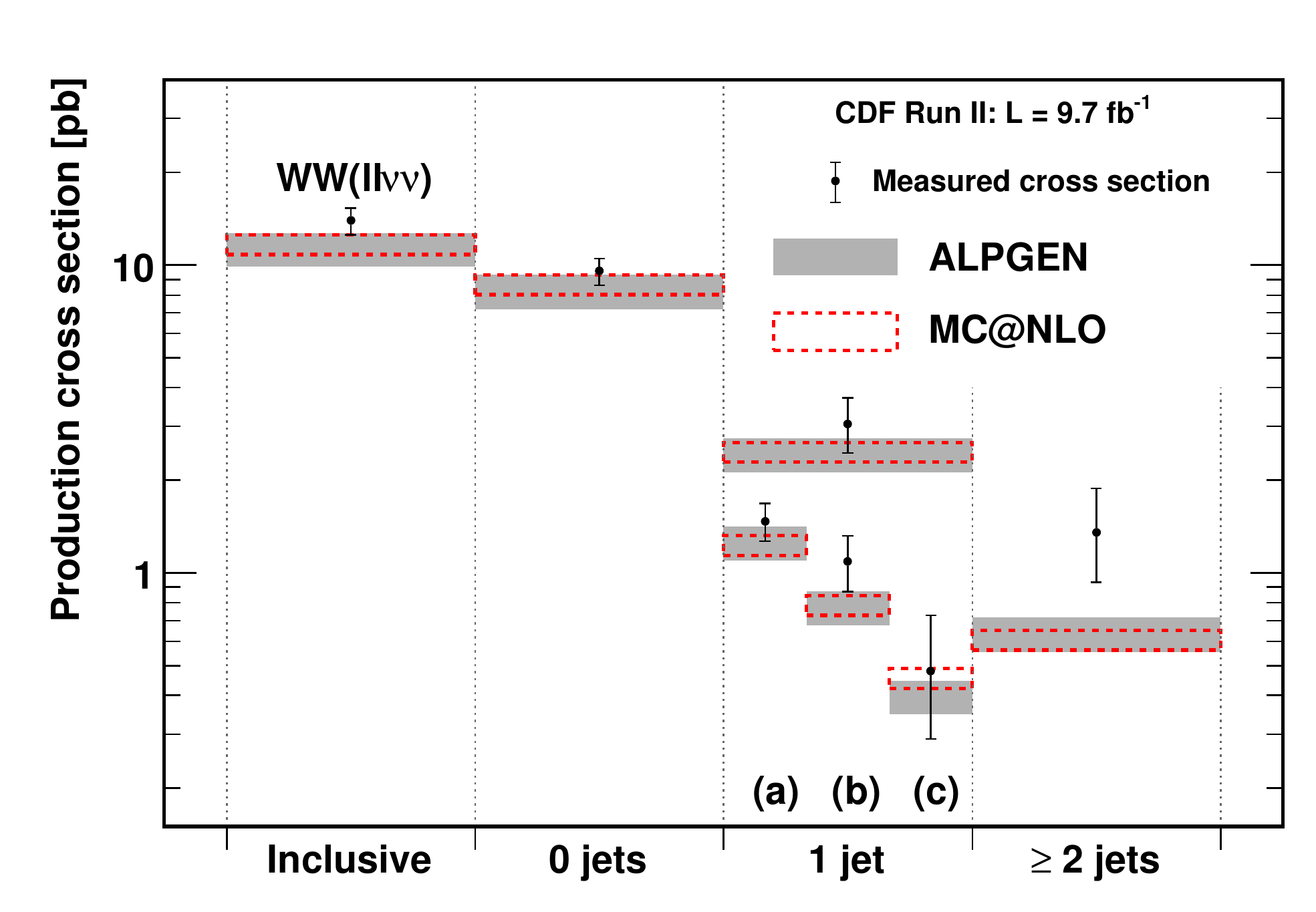}
\hspace{5mm}
\includegraphics[clip,scale=0.3]{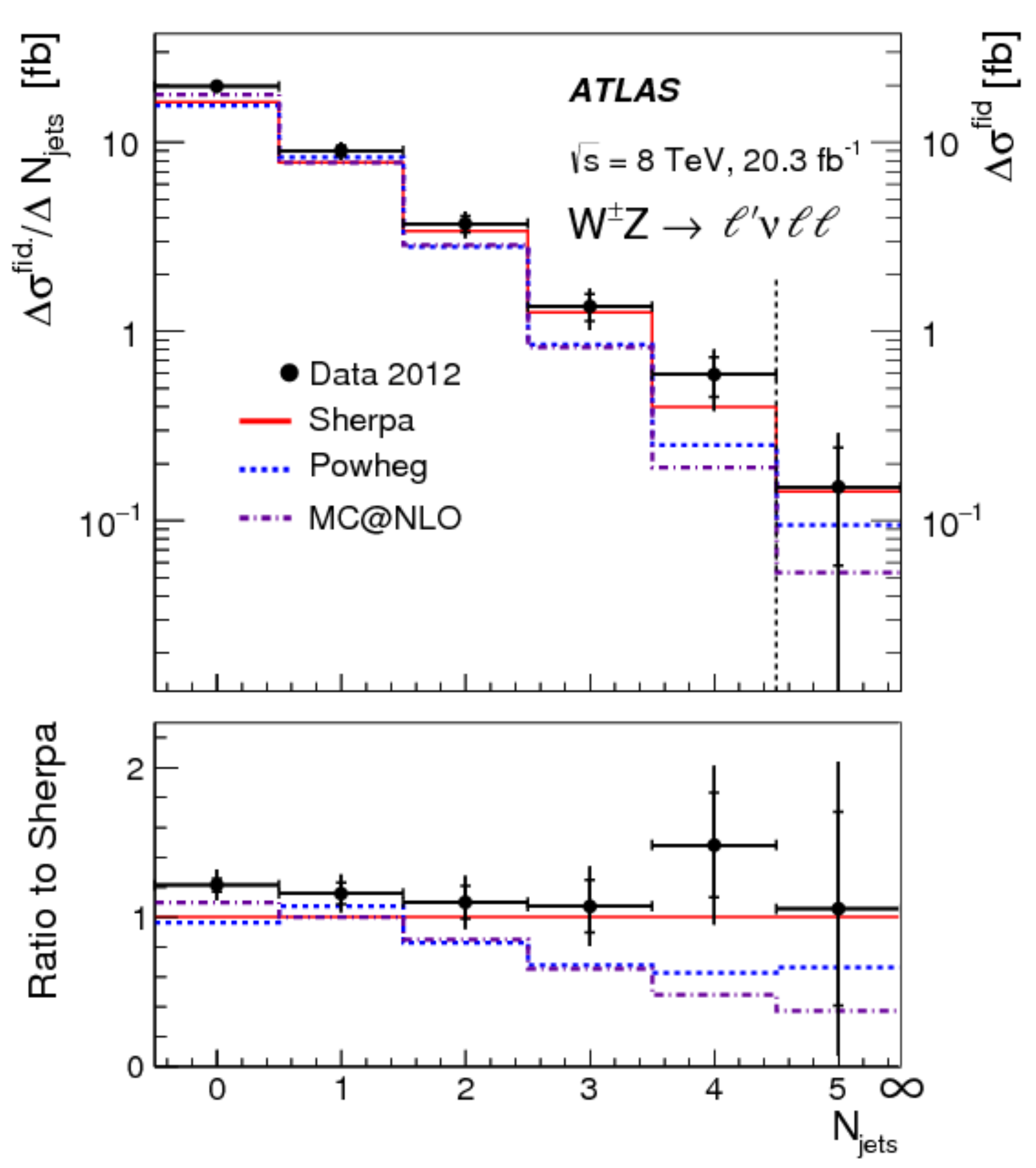}
\caption{
Experimental measurements of di-vector boson production in association with many
jets. On the left CDF's measurement~\cite{WWjCDF} of \WpWm{} production in bins
of jet multiplicity, including an inset for the $E_{\rm T}$ distribution of the
leading jet in inclusive \WpWm$+1$-jet production at the Tevatron. On the right
the ATLAS collaboration show their measurement~\cite{WZATLAS} for $W+Z$ jet
production in bins of jet multiplicity at the LHC, including up to the case of
$WZ+5$ jets.
}
\label{ExpVVjets}
\end{center}
\end{figure*}

At the LHC experiments many studies of inclusive di-vector boson production
have been performed, mainly motivated by their importance for Higgs
phenomenology. Dedicated studies of di-vector bosons measured in bins of jet
multiplicity, however, are scarce. This is due to constraints from dataset
sizes as well as the intrinsic difficulties of measuring multi-lepton
multi-jet signatures. 
Nevertheless, recently the CDF collaboration has measured \WpWm{} plus
multi-jet production at the Tevatron~\cite{WWjCDF} and the ATLAS collaboration
$WZ$ plus many jets at the LHC~\cite{WZATLAS}. In Figure~\ref{ExpVVjets} we
collected results of these two analyses. The impressive advances by ATLAS,
showing results with up to 5 jets in the
final state, is compared to LO (\SHERPA) and NLO predictions (\Powheg{} and
\MCatNLO) combined with parton showers. The Sherpa predictions include matrix elements with the vector-boson pair and up to three hard partons. In contrast,
the \Powheg{} and \MCatNLO{} predictions include hard scattering matrix elements of a
vector-boson pair with at most one associated jet. Such differences explain the
deviations of the theory predictions and motivate further theoretical work for
this signature. 
Presently, inclusive fixed-order predictions beyond two associated jets are not
available for the $WZ$ final state.  We hope to make this next level of
predictions, i.e. $VV$+3-jet results at NLO QCD, available in the near future.
Here we focus on the related signature of \WpWm{} production in association with
up to three jets.

\section{The Calculation}

We employ the \BlackHat{} library~\cite{BlackHatI}, based on unitarity and
on-shell techniques for the computation of the one-loop matrix elements. This
library has been used for many NLO QCD studies in pure-jet production, single-vector-boson-production plus jets and di-photon plus jets (see for
example~\cite{W4jBH,Z4jBH,W5jBH} and references therein). New additions to the
library include: on-shell recursion relations for
tree amplitudes with quarks, gluons and several vector bosons; the
implementation of off-shell tree-level recursions as a cross check; extensions
of the unitarity engine to handle these trees and the
assemblies of loop and tree-level helicity amplitudes into squared matrix
elements. We employ the \SHERPA{} library~\cite{Sherpa} to compute
real-radiation corrections, to integrate over phase space and to produce {\it
ntuple} files~\cite{BHntuples}, which are used to compute general infrared-safe
observables.

We use a leading-color approximation only for the one-loop matrix elements in
\WpWmjjj-jet production. By explicit comparison to lower-point results (\WpWmupjj{} jets) we expect this approximation to be reliable at the percent level. We
have dropped all contributions from closed massive-quark loops and we work with a
diagonal CKM matrix. We decay the $W$ bosons into distinct massless lepton
flavors ($W^+\rightarrow\mu^++\nu_\mu$ and $W^-\rightarrow e^-+\bar\nu_e$).

All results shown have been produced for the LHC with $\sqrt{s}=13$ TeV,
employing the MSTW2008 set of PDFs~\cite{MSTW2008}. We take the strong coupling
$\alpha_s$ provided by the PDF sets consistently at each order of the
perturbative expansion. To set the renormalization
($\mu_r$) and factorization ($\mu_f$) scales, we use a dynamical scale
$\mu=\mu_r=\mu_f=\hat H_{\rm T}/2$, which is half the scalar sum of the
transverse momentum of the partons and leptons in the final state.
The $W$ and $Z$ boson mass
and width are given respectively by $\Gamma_W=2.085$~GeV, $M_W=80.399$~GeV and
$\Gamma_Z=2.4952$~GeV, $M_Z=91.188$~GeV. 

For the lepton sector we impose the following kinematical cuts:
\begin{equation}
\begin{array}{lll}
p_{\rm T}^{e,\mu} > 20 \hbox{ GeV} \,, \hskip 1.5 cm 
&|\eta^{e,\mu}| < 2.4\,, \hskip 1.5 cm 
&\slashed{E}_{\rm T} > 30 \hbox{ GeV}\,,  \hskip 1.5 cm \nonumber\\
p_{\rm T}^{e\mu} > 30 \hbox{ GeV}\,, \hskip 1.5 cm 
&m^{e\mu}>10 \hbox{ GeV}\, , &\, 
\end{array}
\end{equation}
where $p_T^{e\mu}$ and $m^{e\mu}$ represent the transverse momentum and mass of
the electron-muon system respectively. For defining jets we employ the
anti-$k_{\rm T}$ algorithm together with the following cuts:
\begin{equation}
\begin{array}{lll}
p_{\rm T}^{jet} > 30 \hbox{ GeV} \,, \hskip 1.5 cm 
&|\eta^{jet}| < 4.5\,, \hskip 1.5 cm 
&R=0.4\ . 
\end{array}
\end{equation}

\section{Total and Differential Rates}

\begin{figure*}[t]
\begin{center}
\includegraphics[clip,scale=0.45]{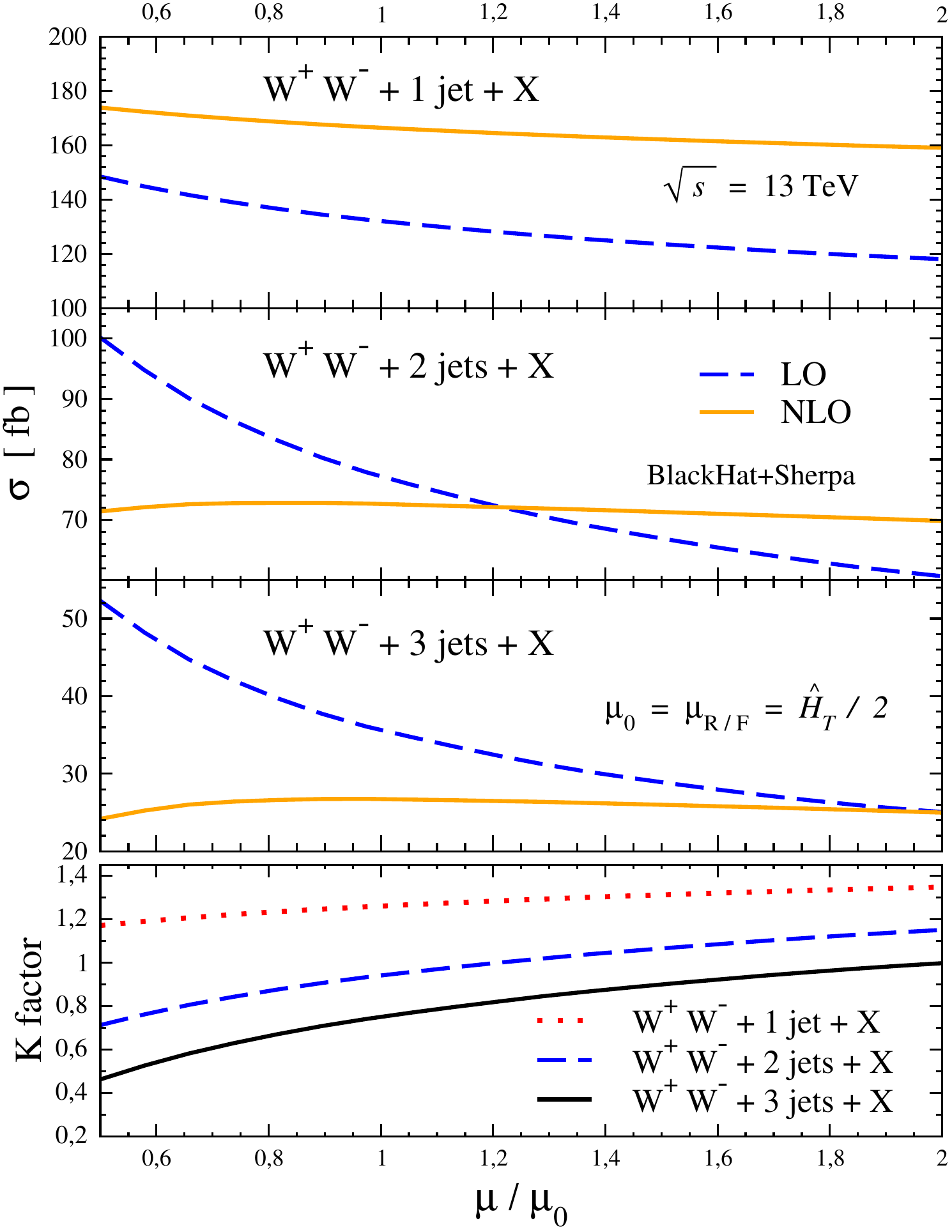}
\caption{
Scale dependence for total cross sections in \WpWm+1,2,3 jets at the LHC. LO
results are shown with the solid (orange) lines and NLO with the dashed (blue)
lines. The bottom panel shows the ratio NLO to LO, so called K factor, for all
multiplicities.
}
\label{xs_scale_dependence}
\end{center}
\end{figure*}

In Figure~\ref{xs_scale_dependence} we show the dependence of the total
inclusive cross sections on the unphysical renormalization and factorization
scales for \WpWm{} production in association with one, two and three jets. The
bottom panel of the figure shows the ratio of NLO cross sections to LO cross
sections, the so called K factors. It can be seen that for all multiplicities a
marked reduction in the spurious scale dependence is achieved. Even more, these
improvements become all the more relevant for larger multiplicities.

\begin{figure*}[t]
\begin{center}
\includegraphics[clip,scale=0.53]{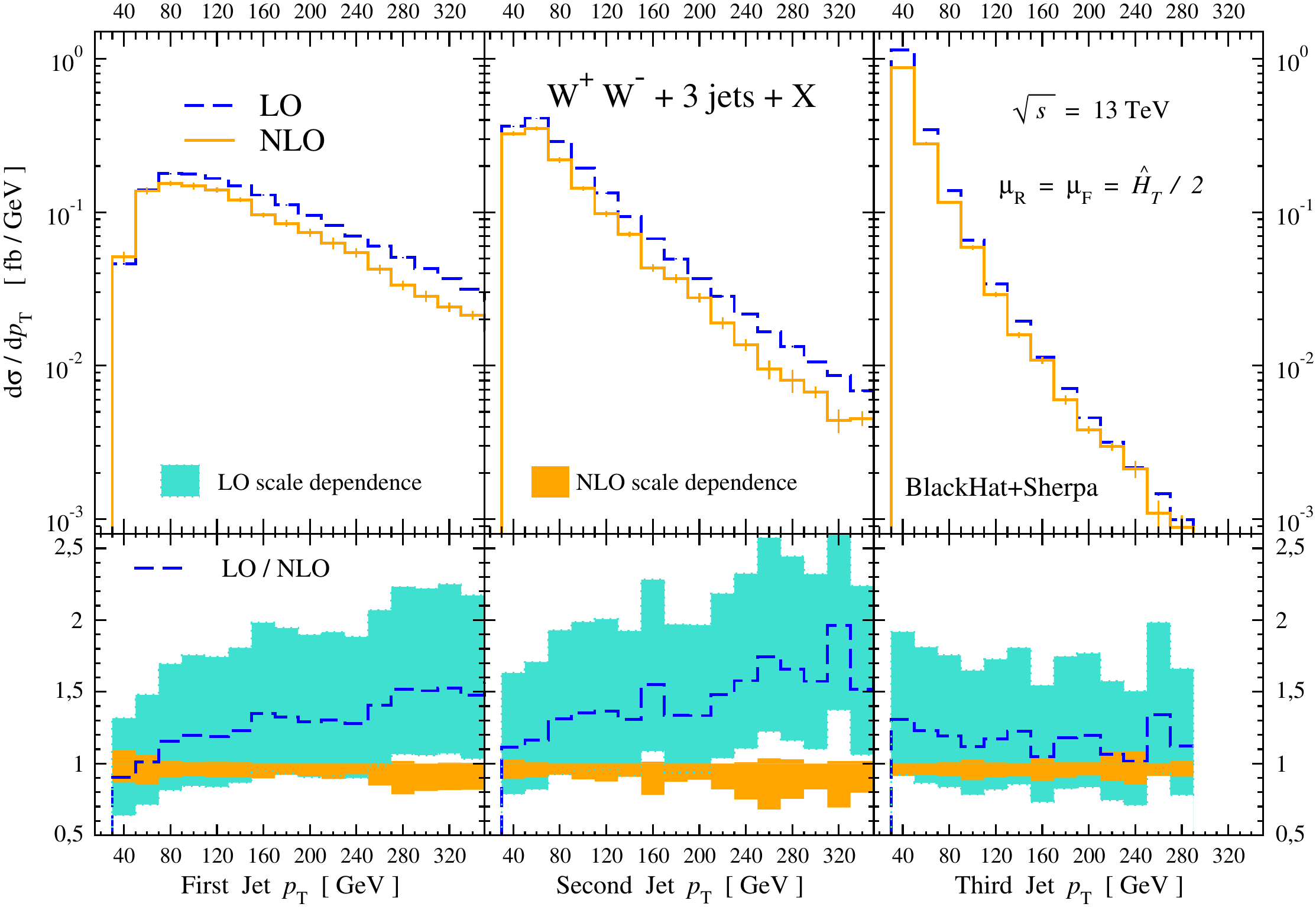}
\caption{A comparison of the $p_T$ distributions of the leading three
jets in \WpWmjjj{}-jet production at the LHC at $\sqrt{s}=13$~TeV.  In
the upper panels the NLO distribution is the solid (orange) histogram
and the LO predictions are shown as dashed (blue) lines.  The thin
vertical line in the center of each bin (where visible) gives its
numerical (Monte Carlo) integration error.  The lower panels show the
LO distribution and the scale-dependence bands 
normalized to the central NLO prediction.
The bands are shaded (orange) for NLO and light-shaded
(cyan) for LO. 
}
\label{WW3ptFigure}
\end{center}
\end{figure*}

We also explore the impact of the quantum corrections over phase space. In
Figure~\ref{WW3ptFigure} we show the jet $p_{\rm T}$ spectra for \WpWmjjj-jet
production. The softer the jet the more steeply their $p_{\rm T}$ distribution
falls. This is very relevant for quantifying the large impact of jet-energy-scale uncertainties on measuring large-jet multiplicity processes. A good feature of our
dynamical scale choice is that the NLO QCD correction do not significantly affect
the shape of the $p_{\rm T}$ distribution of the softest jet. This feature is very
similar to what has been observed in studies of single-vector-boson production
and jets (see for example~\cite{W5jBH}). But not all features of the quantum
corrections can be accounted for by a single scale choice. This is clear for
example by looking at the first and second jet $p_{\rm T}$ distributions. Notice
in general the large reduction of scale dependence at NLO in the bottom panels.

\begin{figure*}[t]
\begin{center}
\includegraphics[clip,scale=0.37]{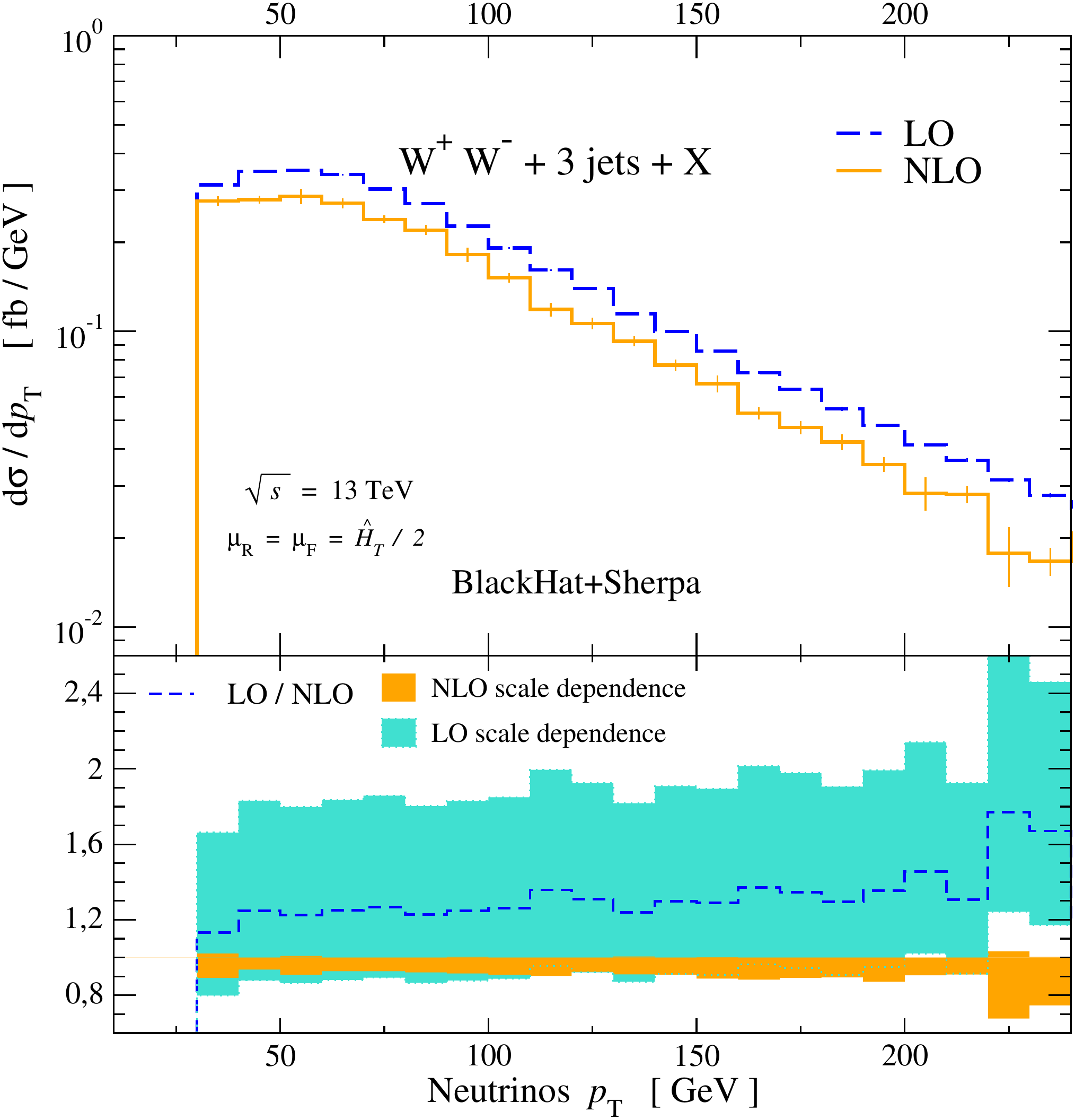}
\caption{
The missing $p_{\rm T}$ distribution for \WpWmjjj-jet production. This
observable is associated to the $p_{\rm T}$ of the $\nu_\mu$-$\bar\nu_e$
neutrino-pair system. Format as in Figure~\protect\ref{WW3ptFigure}.
}
\label{WW3PTmiss}
\end{center}
\end{figure*}

In Figure~\ref{WW3PTmiss} we show the distribution of the $p_{\rm T}$ of
the neutrino-pair system. As neutrinos escape the detector leaving no traces,
they are a source for missing transverse energy. This is a key observable for
the process type that we study. NLO corrections for \WpWmjjj-jet production for
this observable appear stable, particularly for large transverse
momentum; they show only a minor change in the shape of the distribution and LO and NLO scale
dependence bands overlap.

\begin{figure*}[t]
\begin{center}
\includegraphics[clip,scale=0.33]{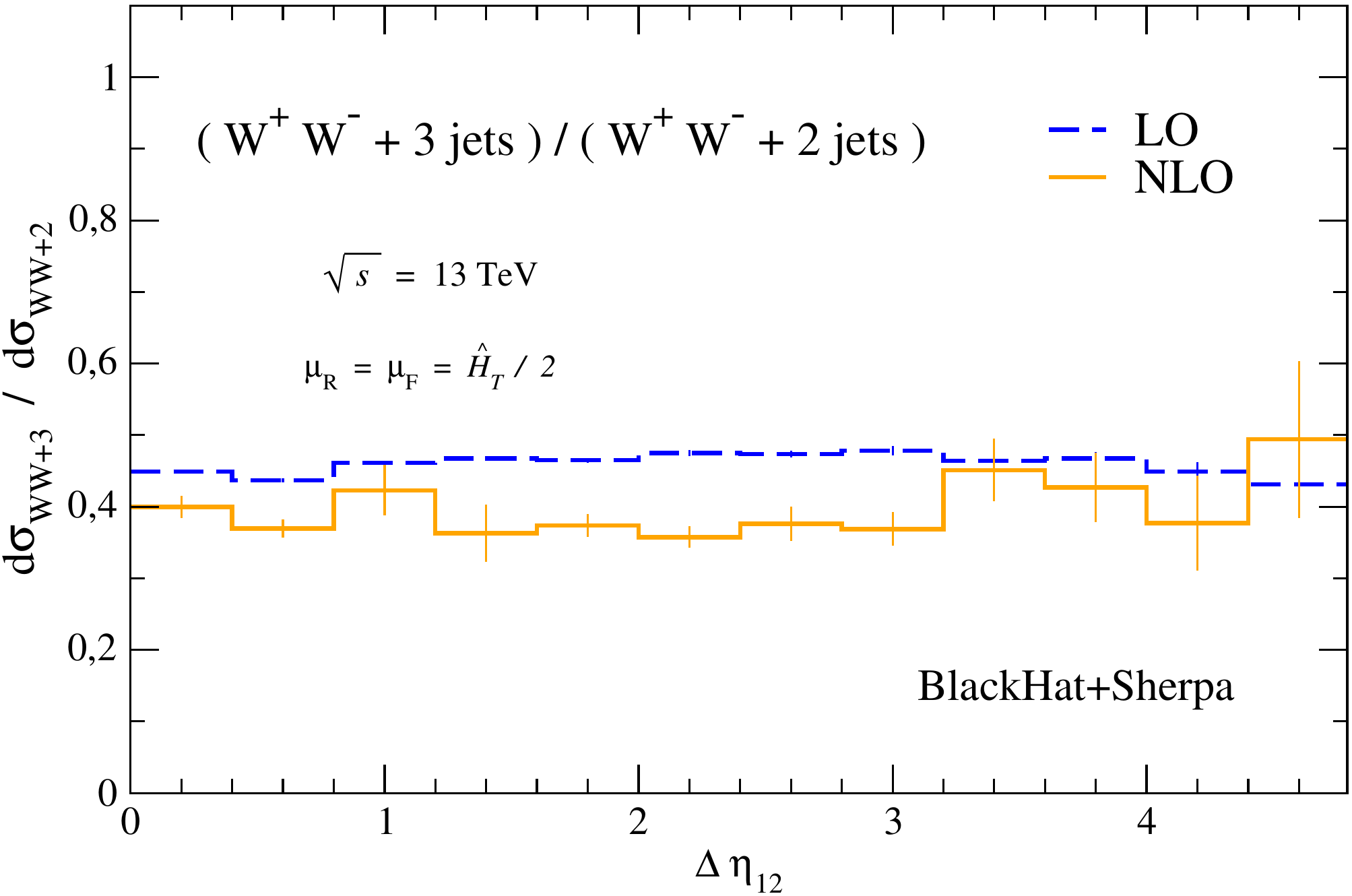}
\hspace{6mm}
\includegraphics[clip,scale=0.33]{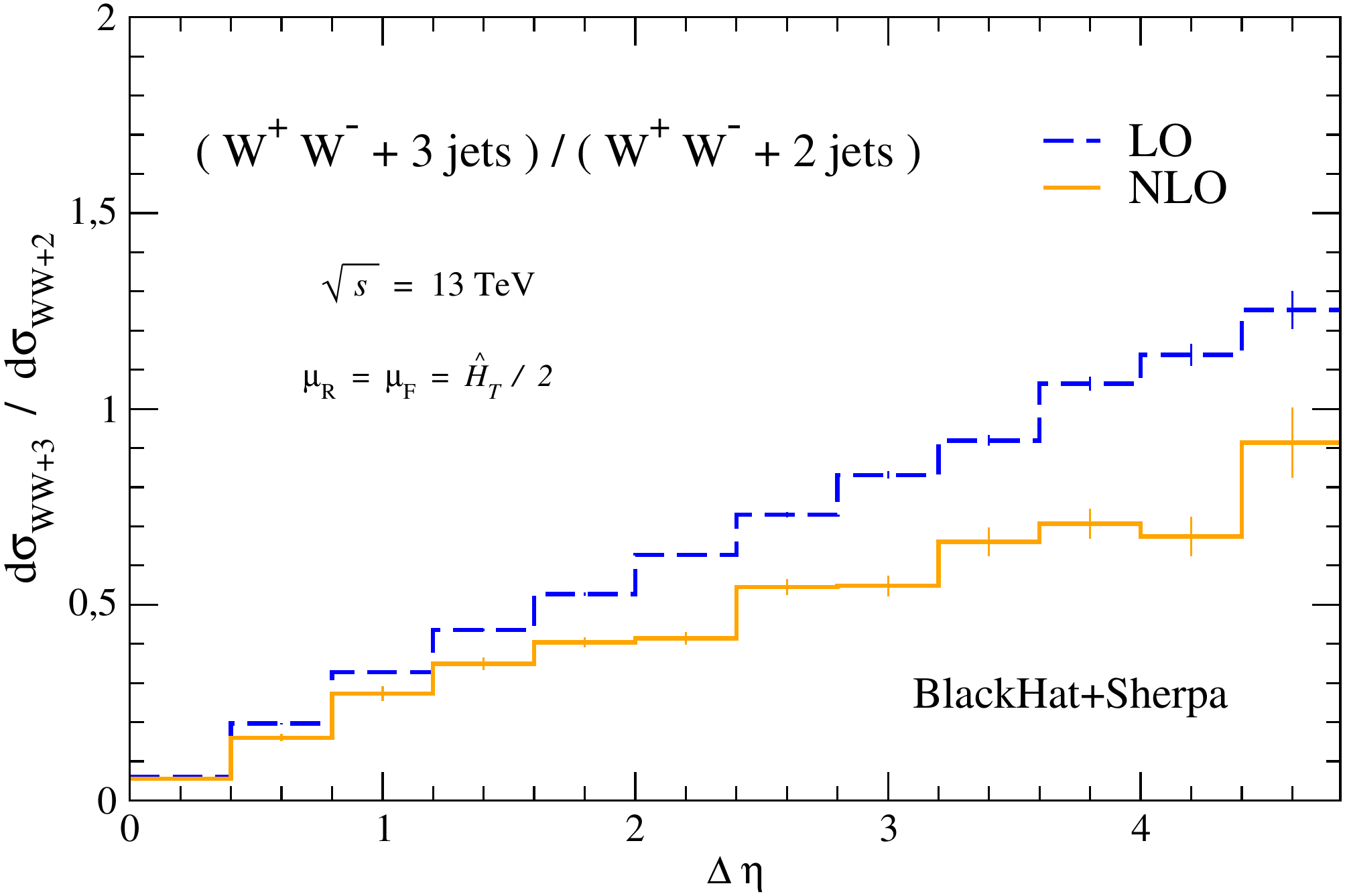}
\caption{
Probability of emission of an extra jet in \WpWmjj-jet production at the LHC. On
the left the tagging jets are chosen by their $p_{\rm T}$, on right the
most-forward and most-backward jets are chosen.
}
\label{RadiationGap}
\end{center}
\end{figure*}

\section{Radiation Gap}

Vector-boson scattering events, $VV\rightarrow VV$, typically produce two
energetic jets ({\it tagging} jets) in the forward and backward parts of the
detectors. These jets are associated to incoming quarks which emit the
scattering vector bosons.  Since the exchange between the interacting quarks is
colorless, the resulting events tend to suppress hard radiation between the
tagging jets. QCD background processes, however, tend to radiate into the gap,
which is exploited in order to disentangle signal from background events.

Having precise predictions for \WpWmjjj-jet production allows to study the
structure of the radiation into the gap for the main irreducible background to
$W^+W^-\rightarrow W^+W^-$ scattering. The idea is that the ratio of
\WpWmjjj{}-jet to \WpWmjj{}-jet rates gives the probability
of finding an extra jet in an inclusive sample for \WpWmjj-jet production.

In Figure~\ref{RadiationGap} we present this observable as a function of
pseudo-rapidity separation between two tagging jets. Two different
configurations are shown: {\it i)} on the left the tagging jets are selected as
the two leading jets organized in $p_{\rm T}$.  {\it ii)} on the
right, the tagging jets are actually taken as the most-forward and
most-backward jets in the event. Interesting patterns of radiation emerge. When
the jets are organized in $p_{\rm T}$ we find a flat radiation pattern as a
function of the jet-pseudo-rapidity separation $\Delta\eta_{12}$. This
behaviour is to be expected and accounts for democratic radiation in all directions.
In this case quantum corrections only very mildly reduce radiation probability
and no change in the shape of the distribution is observed. The case of
radiation into the rapidity gap has more structure.  Phase-space
effects suppress the emission probability in the regime with
$\Delta\eta\rightarrow 0$. On the other hand, the emission probability
increases rapidly for larger values of the rapidity separation $\Delta\eta$. A
marked change in the radiation pattern is induced by the NLO QCD corrections
and it is important to take into account these effects in experimental studies.
We note that the patterns presented are actually very similar to a related
measurement performed by the D0 collaboration~\cite{WjetsRadiationD0} for the
case of a $W$ boson produced in association with jets.

\section{Conclusions}

We have presented NLO QCD results for the production of a \WpWm{} pair in
association with three jets at the LHC. We also computed the cases with zero, one and two
jets which were already given in the literature. We have shown the
considerable reduction in spurious scale sensitivity that is achieved by
including NLO QCD corrections both at the level of total cross sections and for differential
distributions. Having precise \WpWmjjj-jet predictions, and also for
\WpWmjj-jet, allows to study radiation patterns into the gap. This is of key
importance for studies of di-vector boson scattering. We hope to
extend the calculations to other combinations of di-vector bosons, such as
 $WZ$+jet production in the future.

\vspace{6mm}

\noindent {\bf Acknowledgements: } 
The work of F.F.C. is supported by the Alexander von Humboldt Foundation, in the
framework of the Sofja Kovalevskaja Award 2014, endowed by the German Federal
Ministry of Education and Research.  
H.I.'s work is supported by a Marie Sk{\l}odowska-Curie Action
Career-Integration Grant PCIG12-GA-2012-334228 of the European Union.  
This work was performed on the bwUniCluster funded by the Ministry of Science,
Research and the Arts Baden-W\"urttemberg and the Universities of the State of
Baden-W\"urttemberg, Germany, within the framework program bwHP.

\end{document}